\documentclass[twoside]{article}
\usepackage{Proc_NTSE_18}

\usepackage{graphicx}
\usepackage{amssymb}
\newcommand{\be}{\begin{eqnarray}}
\newcommand{\ee}{\end{eqnarray}}
\newcommand{\LCm}{{\scriptscriptstyle -}} %LC supersripts 
\newcommand{\LCp}{{\scriptscriptstyle +}}

\newcommand{\mbf}[1]{\mathbf{#1}}

\newcommand{\bea}{\begin {eqnarray}}
\newcommand{\eea}{\end{eqnarray}}
\newcommand{\ket}[1]{|\,#1\,\rangle}          % Dirac Ket

\newcommand{\bfq}{{\bf q}_{\perp}}

\begin{document}
\thispagestyle{plain}
%Please use the command \publref{myfilename} to print the reference to your proceedings contribution 
%at the bottom of the page where myfilename should be replaced by the name of your LaTeX file
%(e.g., use the command \publref{Johns} if the LaTeX file of your contribution called Johns.tex):
\publref{Mondal}

\begin{center}
{\Large \bf \strut
%Insert the title of your contribution here
Nucleon form factors\\ from basis light front quantization
\strut}\\
\vspace{10mm}
{ \bf 
%Insert the authors here. Use upper indexes a, b, c, etc., to bind authors with their addresses
% as shown below.
Chandan Mondal$^{a}$, Siqi Xu$^{a}$, Jiangshan Lan$^{a,b}$, Xingbo Zhao$^{a,b}$, Yang Li$^{c}$, Henry Lamm$^{d}$ and James P. Vary{$^c$}}
\end{center}

\noindent{% Insert the addresses  here.
\small $^a$\it Institute of Modern Physics, Chinese Academy of Sciences, Lanzhou 730000, China} \\
{% Insert the addresses  here.
\small $^b$\it University of Chinese Academy of Sciences, Beijing 100049, China} \\
{\small $^c$\it Department of Physics and Astronomy, Iowa State University, Ames, IA 50011, USA } \\
{\small $^d$\it Department of Physics, University of Maryland, College Park, Maryland 20742, USA} 
%\\
% {\small $^d$\it The fourth address}

%The next command defines running titles:
\markboth{
%Put here the list of authors that will be displayed in running titles:
C. Mondal, X. Xiqi, J. Lan, {\it et. al.}}
{%Put here the short title of your contribution that will be displayed in running titles:
Nucleon form factors from basis light front quantization} 
%\collaboration{BLFQ Collaboration}
\begin{abstract}
We investigate the electromagnetic form factors of the nucleon in the framework of basis light front quantization. We compute the form factors using the light front wavefunctions obtained by diagonalizing the effective Hamiltonian consisting of the holographic QCD confinement potential, the longitudinal confinement, and a one-gluon exchange interaction with
fixed coupling. The electromagnetic radii of the nucleon are also computed.
\\[\baselineskip] 
\end{abstract}

\section{Introduction}
Electromagnetic form factors are critical to understanding nucleon structure. There are many experiments and theoretical studies on these form factors and they remain a very active field of research. We refer to the articles \cite{Gao,Hyd,Punjabi,Chakrabarti:2013dda,Mondal:2016xpk} for detailed reviews. It is well known that the  matrix element of electromagnetic current for the nucleon requires two form factors namely Dirac and Pauli form factors,
 \be
 J_{had}^\mu(q^2)=\bar{u}(p')\Big(\gamma^\mu F_1(q^2)+\frac{i\sigma^{\mu\nu}q_\nu}{2M}F_2(q^2)\Big)u(p),
 \ee
 where $q^2=(p'-p)^2=-2p'\cdot p+2M^2=-Q^2$ is the square of the momentum transferred to the nucleon and $M$ is the nucleon mass.
 The normalizations of the form factors are given by
 $ F_1^p(0)=1, F_2^p(0)=\kappa_p=1.793$ for the proton and $F_1^n(0)=0, F_2^n(0)=\kappa_n=-1.913$ for the neutron. Cates et al.\cite{Cates} first decomposed the nucleon form factors into their flavor components. 
 Writing the hadronic current as the sum of quark currents one can  decompose the nucleon electromagnetic form factors into flavor dependent form factors.  Neglecting the strange  quark contribution, the hadronic matrix element for electromagnetic current can be  expressed as 
 \be
 J_{had}^\mu(q^2)=\langle N(p')\mid (e_u \bar{u}\gamma^\mu u+ e_d\bar{d}\gamma^\mu d)\mid N(p)\rangle,
 \ee
 where $e_u$ and $e_d$ are the charges of $u$ and $d$ quarks in units of positron charge($e$). Under  the charge  and isospin symmetry $\langle p\mid \bar{u}\gamma^\mu u\mid p\rangle= \langle n\mid \bar{d}\gamma^\mu d\mid n\rangle$, it is straightforward to write down the flavor form factors in term of the nucleon form factors as
\begin{align}
F_i^u(Q^2)&=2F_i^p(Q^2)+F_i^n(Q^2),\nonumber\\
F_i^d(Q^2)&=F_i^p(Q^2)+2F_i^n(Q^2),~~(i=1,2),
\end{align}
with the normalizations $F_1^u(0)=2, F_2^u(0)=\kappa_u$ and $F_1^d(0)=1, F_2^d(0)=\kappa_d$ where the anomalous magnetic moments for the up and the down quarks are $\kappa_u=2\kappa_p+\kappa_n=1.673$ and $\kappa_d=\kappa_p+2\kappa_n=-2.033$.
 It was shown in \cite{Cates} that though the ratio of Pauli and Dirac form factors for the proton $F_2^p/F_1^p \propto 1/Q^2$,  the $Q^2$ dependence is almost constant for the ratio of the quark form factors $F_2/F_1$ for both $u$ and $d$. The Sachs form factors for the nucleon are written in terms of Dirac and Pauli form factors as
 \begin{align}
 G_E^{N}(Q^2)&= F_1^{N}(Q^2)-\frac{Q^2}{4M^2}F_2^{N}(Q^2),\\
 G_M^{N}(Q^2)&= F_1^{N}(Q^2)+F_2^{N}(Q^2),
 \end{align}
and the electromagnetic radii are defined by
 \begin{align}
 \langle r^2_E\rangle^N&=-6 \frac{d G_E^N(Q^2)}{dQ^2}{\Big\vert}_{Q^2=0},\\
  \langle r^2_M\rangle^N&= -\frac{6}{G_M^N(0)} \frac{d G_M^N(Q^2)}{dQ^2}{\Big\vert}_{Q^2=0}.
  \end{align}
  
The basis light front quantization (BLFQ) approach has been developed for solving many-body bound state problems in quantum field theories~\cite{Vary:2009gt,Zhao:2014xaa,Wiecki:2014,Li:2015zda}.
It is a Hamiltonian formalism incorporating the advantages of the light front dynamics \cite{Brodsky:1997de,Hiller:2016itl}. This formalism has been successfully applied to quantum electrodynamics (QED) systems including the electron anomalous magnetic moment~\cite{Zhao:2014xaa} and the strong coupling bound-state positronium problem \cite{Wiecki:2014}. It has also been applied to heavy quarkonia~\cite{Li:2015zda,Li:2017mlw,Lan:2019img} and $B_c$ mesons~\cite{Tang:2018myz} as QCD bound states. Recently, the BLFQ approach using a Hamiltonian that includes the color singlet Nambu-Jona-Lasinio interaction to account for the chiral dynamics has been applied to the light mesons \cite{Jia:2018ary,Lan:2019vui,Lan:2019rba}. In this work, we study the electromagnetic form factors of the nucleon using the light front wavefunctions (LFWFs) obtained by diagonalizing the effective light front Hamiltonian in the constituent valence quark representation with the potential including the light front holographic QCD in the transverse direction \cite{Brodsky:2014yha}, longitudinal confinement \cite{Li:2015zda}, and one-gluon exchange interaction with a fixed coupling in the framework of BLFQ.
%We compare the results obtained in BLFQ formalism with the available experimental data as well as the light-front quark-diquark model \cite{Mondal:2015uha} constructed from the soft-wall AdS/QCD prediction.
%=====================================
\section{Effective light front Hamiltonian}
%=====================================
The structures of the bound states are encoded in the LFWFs which are obtained as the eigenfunctions of the light front Schr\"{o}dinger equation 
\begin{equation}
H_{\mathrm{eff}}\vert \Psi\rangle=M^2\vert \Psi\rangle,\label{eq:LF_Schrodinger}
\end{equation}
where $H_{\mathrm{eff}}$ is the effective Hamiltonian of the system with the mass squared $M^2$ eigenvalue. 
%We consider the effective light front Hamiltonian for the nucleon consisting of the light front kinetic energy with a confining potential in the transverse direction based on light front holographic QCD \cite{Brodsky:2014yha}, as well as a longitudinal confining potential \cite{Li:2015zda}, and the one-gluon exchange interaction with fixed coupling \cite{Li:2015zda}. 
In general, $\vert \Psi\rangle$ is the eigenvector in the Hilbert space spanned by all Fock sectors. In the valence Fock sector, the effective Hamiltonian for the nucleon wavefunctions that we adopt is given by \cite{Xu:2019xhk,Mondal:2020mpv,Mondal:2019yph}
\begin{align}\label{hami}
H_{\rm eff}&=\sum_a \frac{{\vec k}_{a\perp}^2+m_a^2}{x_a}+\frac{1}{2}\sum_{a,b} \left[\kappa_T^4x_ax_b({ \vec r}_{a\perp}-{ \vec r}_{b\perp})^2-\frac{\kappa_L^4}{(m_a+m_b)^2}\partial_{x_a}(x_a x_b\partial_{x_b})\right]\nonumber\\
&+\frac{1}{2}\sum_{a,b} \frac{C_F 4\pi \alpha_s(Q^2_{ab})}{Q^2_{ab}} \bar{u}_{s'_a}(k'_a)\gamma^\mu{u}_{s_a}(k_a)\bar{u}_{s'_b}(k'_b)\gamma^\nu{u}_{s_b}(k_b)d_{\mu\nu},
\end{align}
where $\sum_a x_a=1$, and $\sum_a \mbf{k}_{a\perp}=0$. 
$m_{a/b}$ is the mass of the quark, and $\kappa_L~(\kappa_T)$ is the strength of the longitudinal (transverse) confinement. $\vec \zeta_\perp \equiv \sqrt{x_ax_b} \vec r_\perp$ is the holographic variable \cite{Brodsky:2014yha}, where $\vec r_\perp= { \vec r}_{a\perp}-{ \vec r}_{b\perp}$ is the transverse separation between two quarks, 
$\partial_x f(x, \vec\zeta_\perp) = \partial f(x, \vec \zeta_\perp)/\partial x|_{\vec\zeta}$.
$Q^2_{ab}=-q^2=-(1/2)(k'_a-k_a)^2-(1/2)(k'_b-k_b)^2$ is the average momentum transfer squared, $C_F =-2/3$ is the color
factor. $d_{\mu\nu}$ is the gluon polarization tensor which reduces to the metric tensor $g_{\mu\nu}$ by summing over
the dynamical one-gluon exchange and the instantaneous gluon exchange and $\alpha_s$ is the running coupling which
can be replaced by a constant for simplicity. Note that we use different quark masses in the kinetic energy term and in the one gluon exchange interaction of the effective light front Hamiltonian to simulate the effects of higher Fock components and the other QCD interactions.
Upon diagonalization of the resulting sparse effective Hamiltonian matrix in a chosen basis representation, one obtains the mass spectrum and corresponding wavefunctions of the system. 
%Using the light-front wavefunctions the electromagnetic properties of the nucleon are investigated. 

In BLFQ, Eq. \eqref{eq:LF_Schrodinger} is expressed in a truncated basis representation of the valence Fock space, and the resulting finite-dimensional matrix is diagonalized numerically.
%One then examines the trends in observables as the basis truncation is relaxed to estimate the results in the ``continuum'' limit. 
The choice of basis is arbitrary as long as it is orthogonal and normalized. We choose the two dimensional harmonic oscillator (`2D-HO') basis in the transverse direction and the discretized plane-wave basis in the longitudinal direction \cite{Vary:2009gt,Zhao:2014xaa,Wiecki:2014,Li:2015zda}.
%For each Fock-particle one employs a 2D-HO wave function, $\phi_{nm}(p_\perp)$ , to describe its transverse degrees of freedom and a plane-wave, $e^{-ip^+x^-/2}$, to describe its longitudinal motion. 
Each single-particle basis state can be identified using four quantum numbers, $\bar \alpha = \{k,n,m,\lambda\}$. The longitudinal momentum of the particle is characterized by the first quantum number $k$. In the longitudinal direction $x^\LCm$, we constrain the system to a box of length $2L$, and impose (anti-)~periodic boundary conditions on (fermions) bosons. As a result, the longitudinal momentum $p^\LCp=2\pi k/L$ is discretized, where the dimensionless quantity $k=1, 2, 3,...$ for bosons and $k=\frac{1}{2}, \frac{3}{2}, \frac{5}{2}, ...$ for fermions. The zero mode for bosons is neglected. %The length parameter $L$ should be chosen to cover the longitudinal extent of the system.
In the many-body basis, all basis states are selected to have the same total longitudinal momentum $P^+=\sum_ip_i^+$,
where the sum is over the particles in a particular basis state. One then parameterizes $P^+$ using a dimensionless variable $K=\sum_i k_i$ such that 
$P^+=\frac{2\pi}{L}K$. For a given particle $i$, the longitudinal momentum fraction $x$ is defined as
\begin{equation}
x_i=\frac{p_i^+}{P^+}=\frac{k_i}{K}.
\end{equation}
$K$ determines the ``resolution'' in the longitudinal direction, and thus the resolution on parton distribution functions.
The longitudinal continuum limit corresponds to the limit $L,K \to \infty$.
%In the transverse direction, one employ a 2D Harmonic Oscillator (HO) basis.
The next two quantum numbers, $n$ and $m$, denote radial excitation and angular momentum projection, respectively, of the particle within the 2D-HO basis in the transverse direction. 
%The 2D-HO basis may be defined by two parameters, mass $M$ and frequency $\Omega$. However, we adopt a single HO parameter $b := \sqrt{M \Omega}$, since our transverse modes depend only on $b$ rather than on $M$ and $\Omega$ individually. The state carrying quantum number $n$ and $m$ has HO eigen-energy $E_{n,m}=(2n+|m|+1)\Omega$.
The choice of the 2D HO basis for BLFQ is made because the HO potential
is a confining potential, and therefore its wavefunctions should form an ideal basis for systems subject to QCD confinement. Since we assume harmonic confinement in the transverse direction, these transverse basis states are also computationally convenient.

In order to numerically diagonalize $H_{\rm eff}$, the infinite dimensional basis must be truncated down to a finite dimension. In BLFQ, two levels of truncation scheme are implemented.
First, the number of Fock sectors in the basis is restricted. This truncation will be based on physical as well as practical considerations.
For instance, the nucleon is expected to be fairly well described by the lowest few sectors.
For example, the nucleon state can be expressed schematically as
\be
\ket{N}_{\hbox{\scriptsize phys}}=a\ket{qqq}+b\ket{qqqg}+c\ket{qqqq\bar{q}}+\cdots.
\ee
In this work, we limit ourselves to only the leading Fock sector $\ket{qqq}$. 
%We explicitly assume that higher Fock sectors give insignificant contributions to the low-lying eigenstates. %We do not make any attempt here to examine the limit of increasing the number of Fock sectors.

Second, within each Fock-sector, further truncation is still needed to reduce the basis to a finite dimension. 
%As mentioned, we impose (anti-)~periodic boundary conditions on (fermions) bosons in a longitudinal box with length $2L$. Consequently, the longitudinal momentum $p^\LCp$ of single particles can only take discrete values. 
We introduce a truncation parameter $K_{\rm max}$ on the longitudinal direction such that, $\sum_l k_l \le K_{\rm max}$, where $k_l$ is the longitudinal momentum quantum number of $l$-th particle in the basis state. Note that systems with larger $K_{\rm max}$ have simultaneously higher ultra-violet (UV) and lower infra-red (IR) cutoffs in the longitudinal direction. In the transverse direction, we require the total transverse quantum number $N_\alpha=\sum_l (2n_l+| m_l |+1)$ for multi-particle basis state $\ket{\alpha}$ satisfies $N_\alpha \le N_\text{max}$, where $N_\text{max}$ is a chosen truncation parameter. The transverse continuum limit corresponds to $N_{\max}\to\infty$. The 2D-HO basis may be defined by two parameters, mass $M$ and frequency $\Omega$. We adopt a single HO parameter $b= \sqrt{M \Omega}$ , since our transverse modes depend only on $b$ rather than on $M$ and $\Omega$ individually. Here, we choose the value of $b=0.45$ GeV, the same as the confining strength $\kappa_L(\kappa_T)$. $N_{\rm max}$ and $b$ define both the transverse IR and UV regulator in BLFQ.
%The state carrying quantum number $n$ and $m$ has HO eigen-energy $E_{n,m}=(2n+|m|+1)\Omega$. 
In addition, our many
body states have well defined values of the total angular momentum projection
$
M_J=\sum_i\left(m_i+\lambda_i\right),
$
where $\lambda$ is the fourth quantum number which corresponds the helicity of the particle.
\begin{figure}[htp]
\begin{center}
(a)\includegraphics[width=0.7\textwidth]{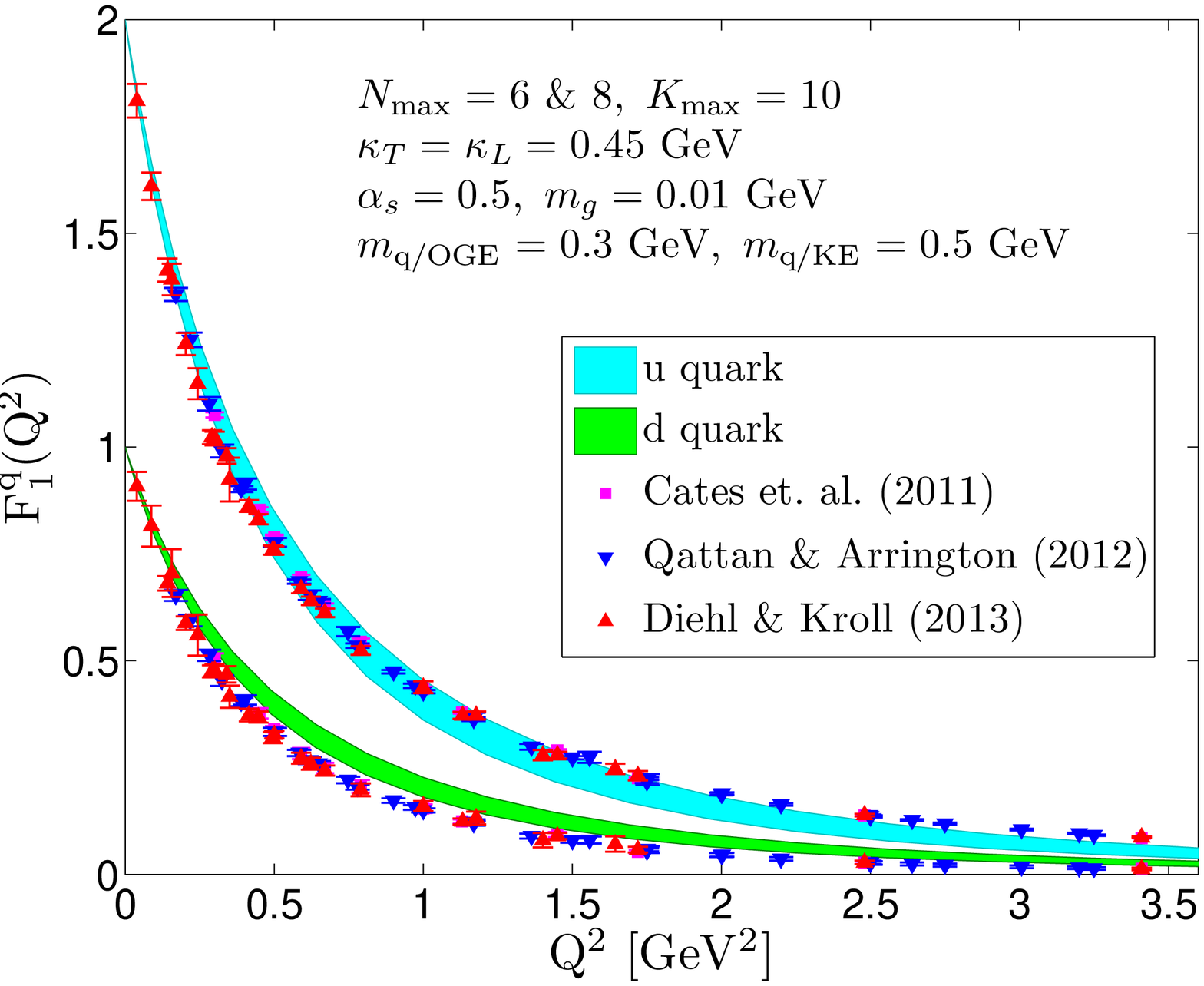}
(b)\includegraphics[width=0.7\textwidth]{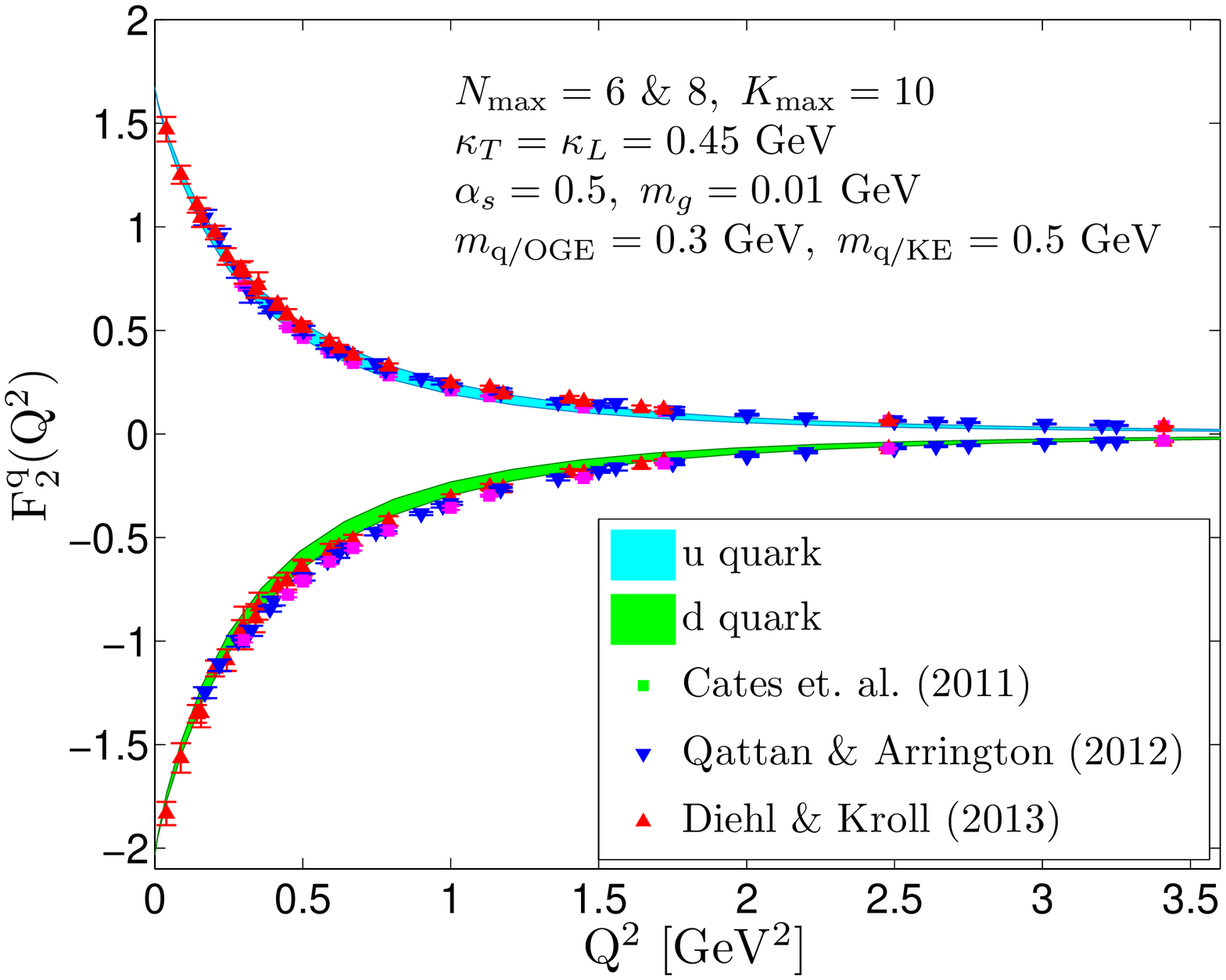}
\caption{(Color online)~BLFQ results for (a) the Dirac and, (b) the Pauli form factors of $u$ and $d$ quarks with confining strength, $\kappa_L=\kappa_T=0.45$ GeV and fixed coupling $\alpha_s=0.5$. The quark mass in the kinetic energy term is $m_{\rm q/KE}=0.5$ GeV, whereas the quark mass in one gluon exchange interaction is $m_{\rm q/OGE}=0.3$ GeV. The bands correspond the range for $N_{\rm max}=6-8$ with $K_{\rm max}=10$. We choose the value of HO parameter b same as  $\kappa_L(\kappa_T)$ i.e. $b=0.45$ GeV. $m_g (=0.01~\rm GeV)$ is a small gluon mass regulator used for numerical convenience.  The experimental data are taken from \cite{Cates,Qattan:2012zf,diehl13}.
}
\label{fig1}
\end{center}
\end{figure}
\begin{figure}[htp]
\begin{center}
\includegraphics[width=0.7\textwidth]{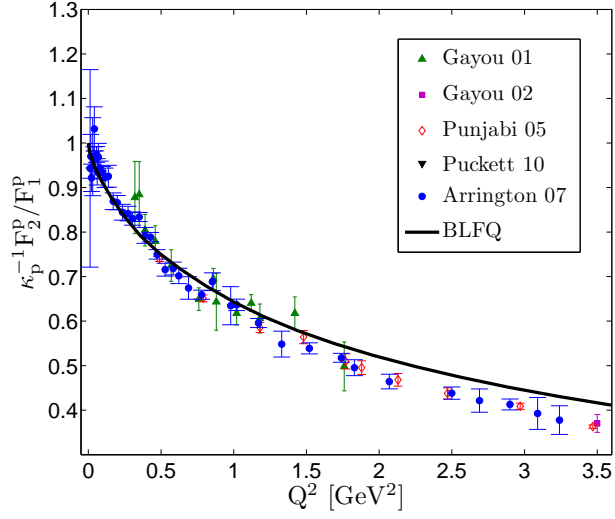}
\caption{(Color online)~The ratio of Pauli and Dirac form factors of the proton with the same parameters as mentioned in Fig. \ref{fig1} and with basis truncation $N_{\rm max}=8$ and $K_{\rm max}=10$. The ratio is divided by $\kappa_p$. The experimental  data are taken from Refs. \cite{Gay1,Gay2,Arr,Pun,Puck}.
}
\label{fig2}
\end{center}
\end{figure}
\begin{figure}[htp]
\begin{center}
(a)\includegraphics[width=0.7\textwidth]{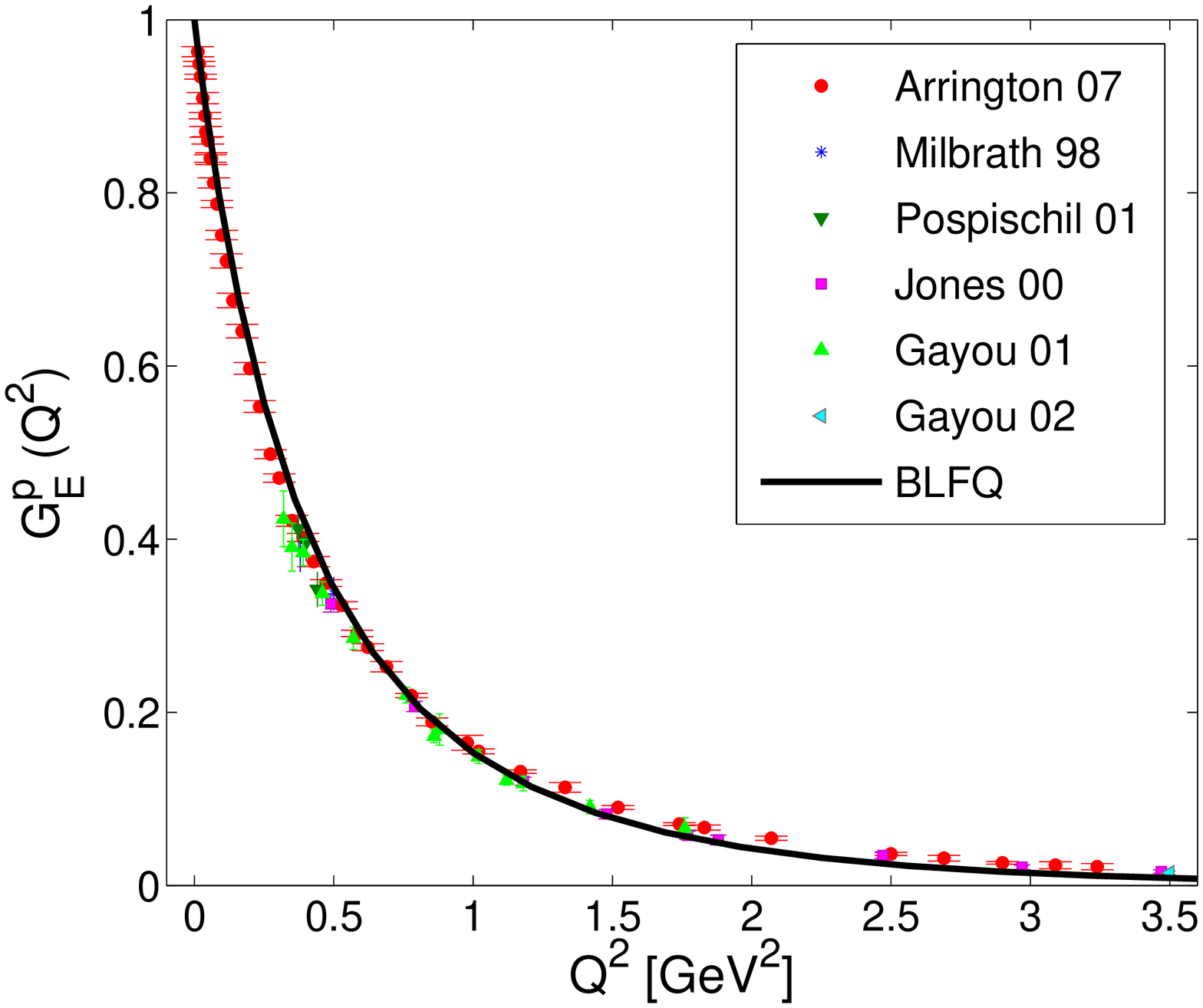}
(b)\includegraphics[width=0.7\textwidth]{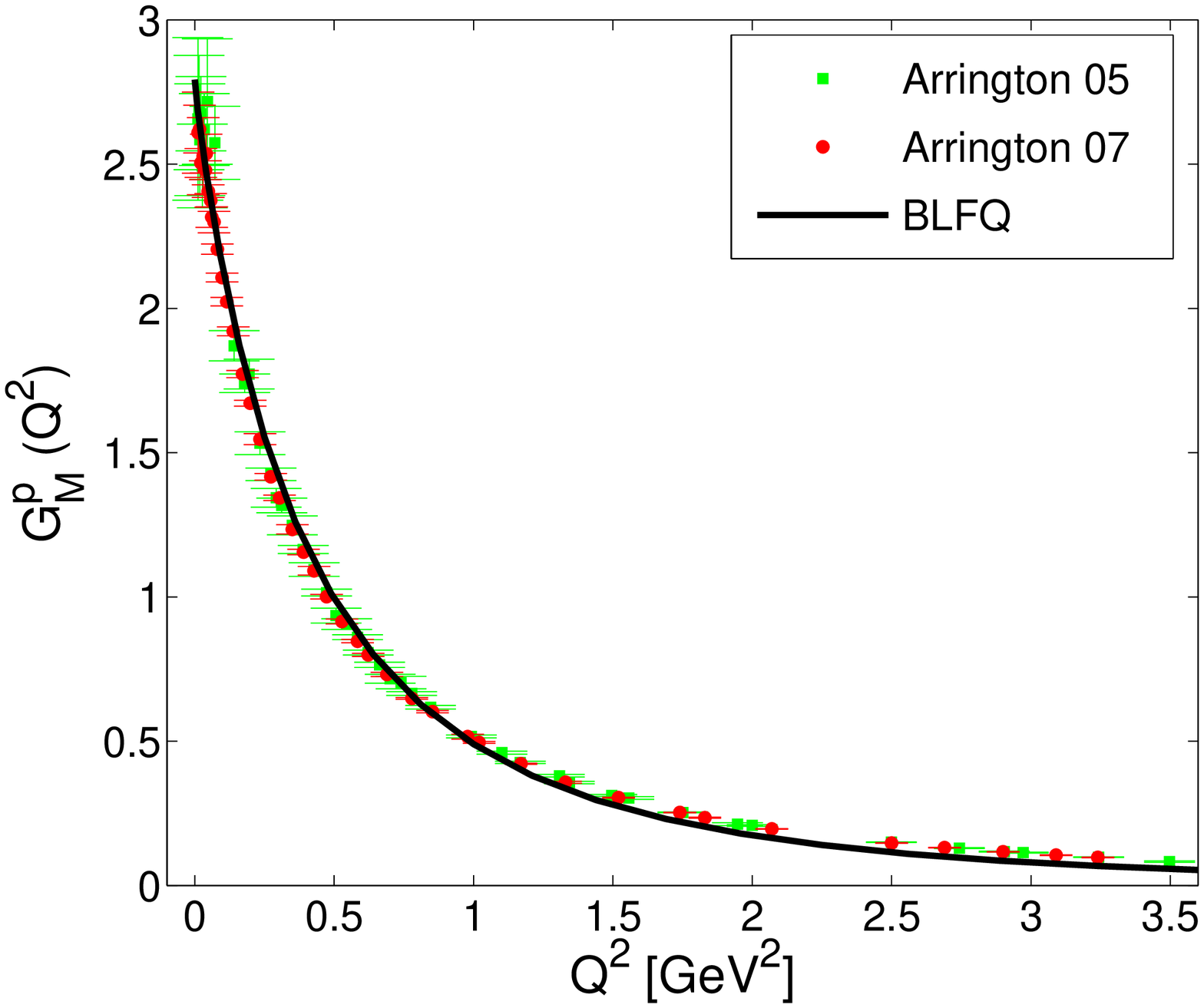}
\caption{(Color online)~(a)BLFQ results for the Sachs form factors (a) $G_E(Q^2)$, and (b) $G_M(Q^2)$ of the proton with the same parameters as mentioned in Fig. \ref{fig1} and with basis truncation $N_{\rm max}=8$ and $K_{\rm max}=10$. The experimental data are taken from Refs.\cite{Gay1,Gay2,Arr,Milbrath,Posp,Jones} and \cite{Arr,Arr2}.
}
\label{fig3}
\end{center}
\end{figure}
%====================
\begin{figure}[htp]
\begin{center}
(a)\includegraphics[width=0.7\textwidth]{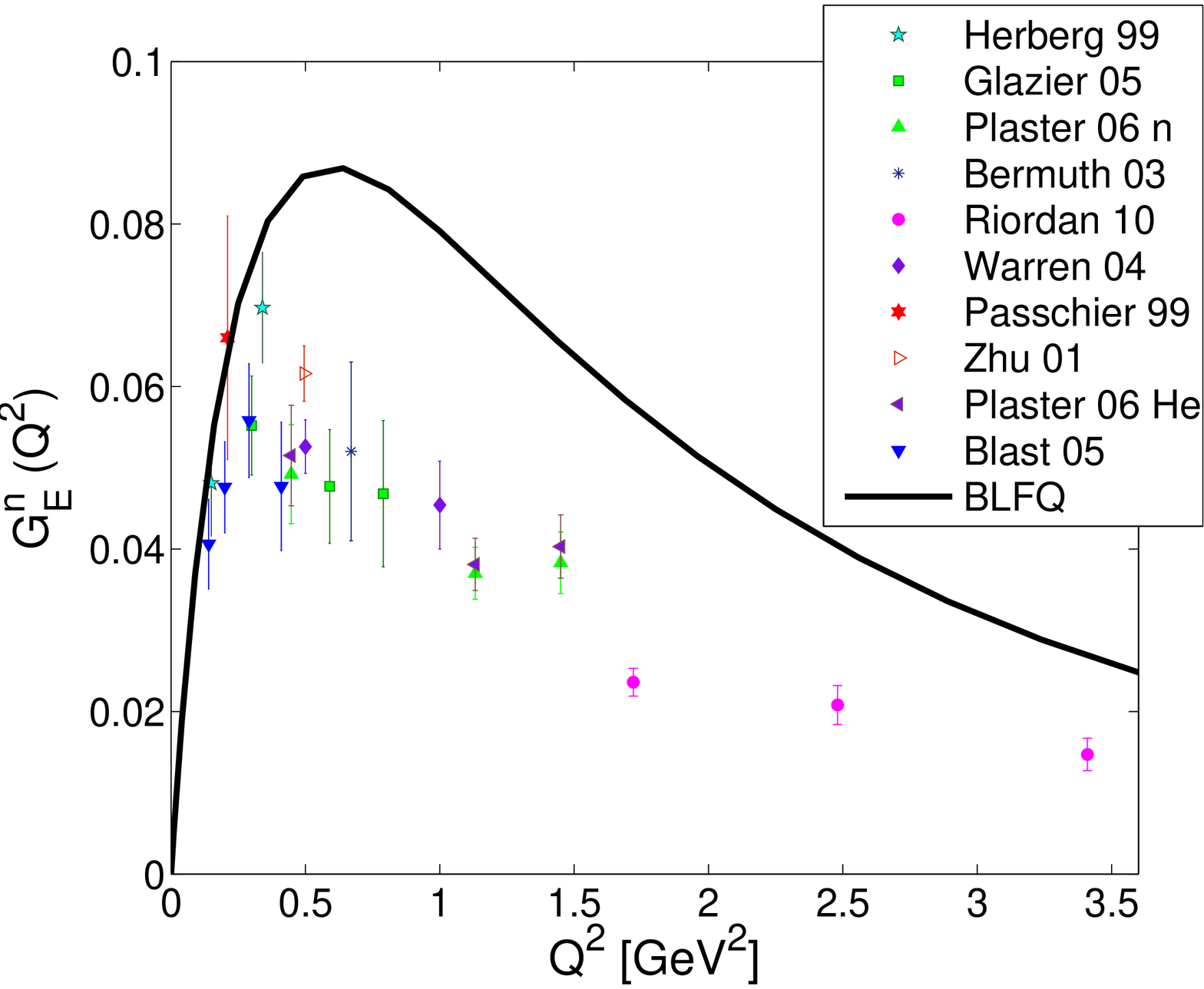}
(b)\includegraphics[width=0.7\textwidth]{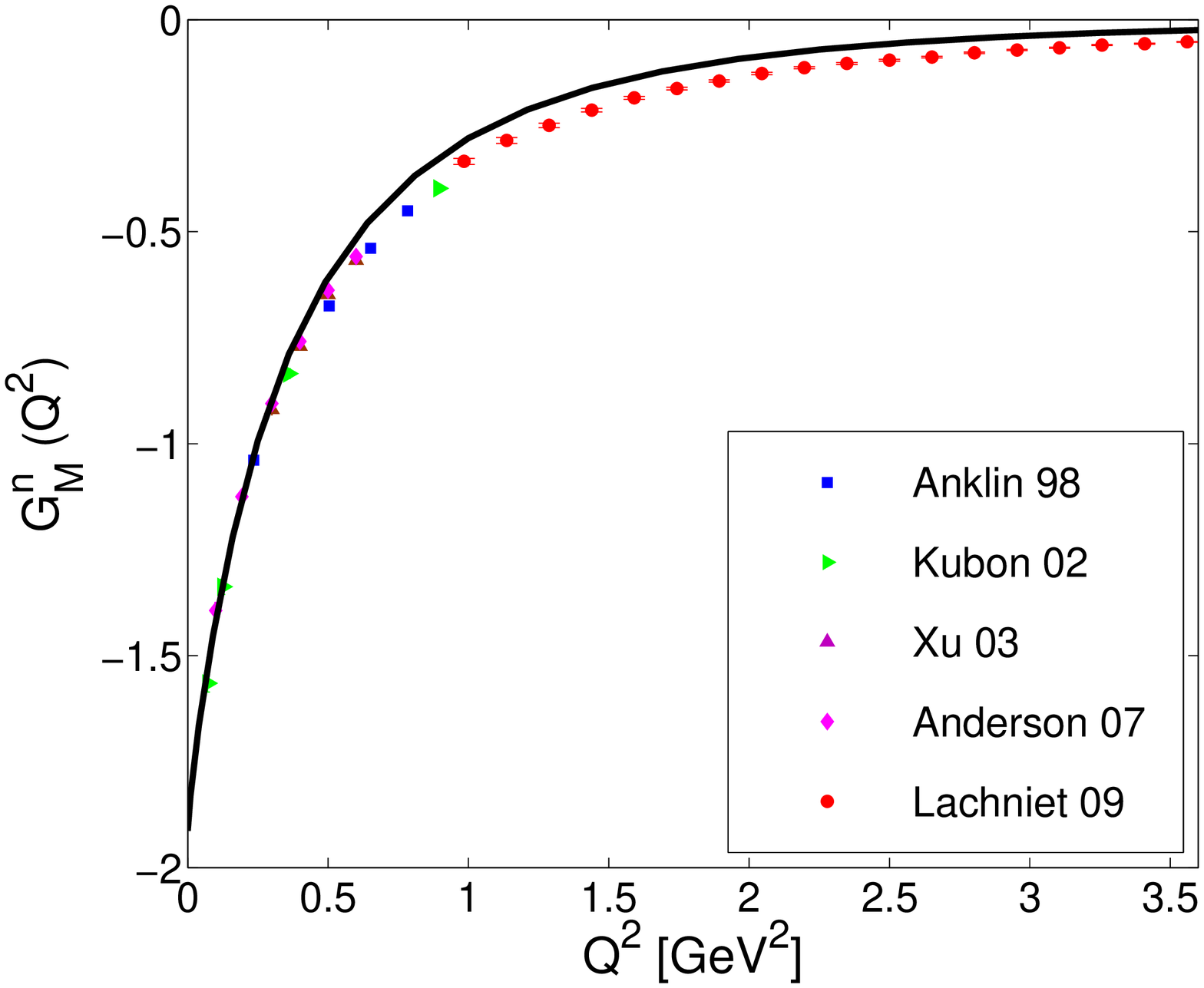}
\caption{(Color online)~(a)BLFQ results for the Sachs form factors (a) $G_E(Q^2)$, and (b) $G_M(Q^2)$ of the neutron with the same parameters as mentioned in Fig. \ref{fig1} and with basis truncation $N_{\rm max}=8$ and $K_{\rm max}=10$. The experimental data are taken from Refs.\cite{Arr,Milbrath,Posp,Jones,Gay1,Gay2} and \cite{Anklin,Kubon,Xu,Anderson,Lachniet}.
}
\label{fig4}
\end{center}
\end{figure}
%====================
%=============================================
\section{Electromagnetic form factors in BLFQ}
%=============================================
In the light front formalism for a spin $\frac{1}{2}$ composite system the Dirac and Pauli form factors $F_1(q^2)$ and $F_2(q^2)$ are identified with the helicity-conserving and helicity-flip matrix elements of the $J^+$ current \cite{BD}
\be
\langle P+q, \uparrow|\frac{J^+(0)}{2P^+}|P, \uparrow\rangle &=&F_1(q^2),\\
\langle P+q, \uparrow|\frac{J^+(0)}{2P^+}|P, \downarrow\rangle &=&-(q^1-iq^2)\frac{F_2(q^2)}{2M},
\ee
where $M$ is the nucleon mass and the arrow indicates the helicity
of the nucleon.
The physical nucleon state with momentum  $P$ can be expanded in terms of multi-particle light front wavefunctions \cite{BDH}:
\begin{align}
\mid P,S_z \rangle &= \sum_{n} \int \prod_{i=1}^n {dx_i d^2 k_{\perp i} \over
\sqrt{x_i} 16 \pi^3} 16 \pi^3 \delta\left(1- \sum_{i=1}^n x_i\right) \delta^2
\left(\sum_{i=1}^n k_{\perp i}\right)\nonumber\\
&\times \psi^{S_z}_n(x_i, k_{\perp i}, \lambda_i)
\mid n, x_i P^+, x_i P_\perp + k_{\perp i}, \lambda_i \rangle;
\end{align}
% \end{widetext}
here $x_i=k_i^+/P^+$   and $k_{\perp i}$ 
represent the relative transverse momentum of the $i$-th constituent and $n$ is the number of particles in a Fock state. The
physical transverse momenta are $p_{\perp i} = x_i P_\perp + k_{\perp i}$. $\lambda_i$ and $S_z$ are the light-cone helicities of the quark and nucleon, respectively. The boost invariant light-front wave functions 
$\psi_n$ depend only on $x_i$ and  $k_{\perp i}$  and are independent of the total momentum of the state $P^+$ and $P_\perp$. 
In the overlap representation, the electromagnetic form factors are then expressed as 
\begin{align}
F_1^q(q^2)&= 
 \sum_{n,\lambda_i} \int \prod_{i=1}^n {dx_i d^2
k_{\perp i}\over 16 \pi^3} 16 \pi^3 \delta\left(1-\sum_j x_j\right)
\delta^2\left(\sum_{j=1}^n k_{\perp j}\right) \nonumber\\& \times \psi_n^{\uparrow*}(x_i',
{k'}_{\perp i},\lambda_i) \psi_n^{\uparrow}(x_i,
k_{\perp i},\lambda_i) ;    \\
\frac{-(q^1-iq^2)}{2 M}F_2^q(q^2)&= 
 \sum_{n,\lambda_i} \int \prod_{i=1}^n {dx_i d^2
k_{\perp i}\over 16 \pi^3} 16 \pi^3 \delta\left(1-\sum_j x_j\right)
\delta^2\left(\sum_{j=1}^n k_{\perp j}\right) \nonumber\\& \times \psi_n^{\uparrow*}(x_i',
{k'}_{\perp i},\lambda_i) \psi_n^{\downarrow}(x_i,
k_{\perp i},\lambda_i) ; 
\end{align}
where  for the struck parton %\Xingbo{constituent} 
 ${x'}_1=x_1; 
~{k'}_{\perp 1}=k_{\perp 1}+(1-x_1) q_\perp$ and  ${x'}_i={x_i}; ~{k'}_{\perp i}=k_{\perp i}-{x_i} q_\perp$ for  the spectators ($i=2,....n$). We consider the frame where $q=(0,0,\bfq)$,  thus $Q^2=-q^2={\bf q}_\perp^2$. 
Since we restrict ourselves to the leading Fock sector, the nucleon basis state can be written as 
\be
\ket{N_{\hbox{\scriptsize phys}}^{S_z}}=\ket{k_{q_1},n_{q_1},m_{q_1},\lambda_{q_1}}\otimes\ket{k_{q_2},n_{q_2},m_{q_2},\lambda_{q_2}}\otimes\ket{k_{q_3},n_{q_3},m_{q_3},\lambda_{q_3}}. \label{basis}
\ee
%In terms of the BLFQ amplitude of the nucleon state, $\braket{N_\text{phys}^{S_z}}{k_{q_i},n_{q_i},m_{q_i},\lambda_{q_i}}$, we express the Dirac and the Pauli form factors as
%\begin{align}
%    \label{F1}
%&F_1^q\left(Q^2\right)=\sum_{\rm all} \braket{{N'}^\up_\text{phys}}{k_{q_i},n_{q_i},m_{q_i},\lambda_{q_i}}\braket{k_{q_i},n_{q_i},m_{q_i},\lambda_{q_i}}{N^\up_\text{phys}},    \\
%    \label{eq:GPD_E_BLFQ_unrenormalized}
%&\frac{q^1-iq^2}{2 M}F_2^q(Q^2)=\sum_{\rm all} \braket{{N'}^\up_\text{phys}}{k_{q_i},n_{q_i},m_{q_i},\lambda_{q_i}}\braket{k_{q_i},n_{q_i},m_{q_i},\lambda_{q_i}}{N^\down_\text{phys}};,
%\end{align}
%where the summation is over all the quantum numbers of all the constituents.
We obtain the light front wavefunctions numerically by diagonalizing the effective Hamiltonian given in Eq.(\ref{hami}) with the basis representation given by Eq. (\ref{basis}).
Using the resulting light front wavefunctions $\psi_n$, we evaluate the electromagnetic form factors of the nucleon. The parameters are tuned to fit the electromagnetic properties of the nucleons.
Following the convention of \cite{Mondal:2015uha}, we fix the normalizations  of  the Dirac and the Pauli form factors as
\be
F_1^q(Q^2)=n_q\frac{F_{1}^{\rm(BLFQ)q}(Q^2)}{F_{1}^{\rm(BLFQ)q}(0)},~~~~~~~~~F_2^q(Q^2)=\kappa_q\frac{F_{2}^{\rm(BLFQ)q}(Q^2)}{F_{2}^{\rm(BLFQ)q}(0)},\label{D_P_FF}
\ee
so that $F_1^q(0)=n_q$  and $F_2^q(0)=\kappa_q$ where $n_u=2,~n_d=1$ and the anomalous magnetic moments for the $u$ and $d$ quarks are $\kappa_u=1.673$ and $\kappa_d=-2.033$.  The advantage of the modified formulae in Eq.(\ref{D_P_FF}) is that, irrespective of the values of the parameters, the normalization conditions for the form factors are automatically satisfied.

In Fig. \ref{fig1}, we show the $Q^2$ dependence of the Dirac and the Pauli form factors of $u$ and $d$ quark. We set the confining strength, $\kappa_L=\kappa_T=0.45$ GeV in both the longitudinal and transverse confinements and the coupling constant $\alpha_s=0.5$. The bands represent the range of our results due to increasing the basis from  $N_{\rm max}=6$ to $N_{\rm max}=8$ with $K_{\rm max}=10$. 
%The lower and upper lines in each band correspond the result for $N_{\rm max}=6$ and $N_{\rm max}=8$ respectively. 
We use different quark masses i.e. in the kinetic energy term, $m_{\rm q/KE}=0.5$ GeV and in the one gluon exchange interaction, $m_{\rm q/OGE}=0.3$ GeV in order to minimize the effect of higher Fock component and the other QCD interactions.  Fig. \ref{fig1} shows that the BLFQ results for the flavor Pauli form factors are in reasonable agreement with the experimental data. The Dirac form factor for the $u$ quark is also in reasonable agreement with the data. However, the theoretical $d$ quark form factor is somewhat over estimated compared to the data. 

The nucleon form factors can be obtained from the flavor dependent form factors.
The ratio of Pauli and Dirac form factors of the proton for $N_{\rm max}=8$ and $K_{\rm max}=10$ is shown in Fig. \ref{fig2}. We find that at low $Q^2$ our result agrees well with the experimental data. The Sachs form factors for the proton are presented in Fig. \ref{fig3} where we find a good agreement between theory and experiment. In Fig. \ref{fig4}, we show the Sachs form factors for the neutron. Our results for the neutron magnetic form factor is in reasonable agreement with experimental data, however, for the charge form factor is over estimated compared to the data. The deviations of the neutron charge form factor from the experimental data can be attributed to the fact that the $d$ quark form factor $F_1^d$ does not have the correct behavior in this model.  From the Sachs form factors we also compute the electromagnetic radii of the nucleons. We quote the radii in Table \ref{table}, the experimental values are taken from the  Ref. \cite{pdg}. Here again, we find reasonable agreement with experiment.
%%%%%%%%%%%%%%%%%%%%%%%%%%%%%%%%%%%%%%%%%%%%%%%%%%%%%%%%%%%%%%%%%%%%%%%
%\begin{widetext}   
\begin{table}[ht]
\centering % used for centering table 
\begin{tabular}{ |c| c| c| } % centered columns (3 columns) 
\hline
Quantity&~~~BLFQ~~~& ~~~Measured data\cite{pdg}~~~  \\ [0.5ex] % inserts table 
%heading 
\hline % inserts single horizontal line 
$r_E^p$ & $0.804$ fm  & $0.877 \pm 0.005$  fm\\ 
$r_M^p$ & $0.917$ fm & $0.777\pm0.016$ fm \\ 
\hline
$\langle r_E^2\rangle^n$ & ${-}0.1214$ fm$^2$ & $-0.1161 \pm 0.0022$ fm$^2$ \\

$r_M^n$ & $1.007$ fm  &  $0.862^{+0.009}_{-0.008} $ fm\\
\hline %inserts single line 
\end{tabular} 
\caption{Electromagnetic radii of the nucleons.}% title of Table 
\label{table} % is used to refer this table in the text 
\end{table} 
%\end{widetext}
%%%%%%%%%%%%%%%%%%%%%%%%%%%%%%%%%%%%%%%%%%%%%%%%%%%%%%%%%%%%%%%%%%%%%%%
%====================
\section{Conclusions}
%====================
The electromagnetic form factors for the nucleon and their flavor decomposition have been presented using the BLFQ approach. The form factors have been evaluated from the overlaps of the light front wavefunctions which were obtained by diagonalizing the effective Hamiltonian. In our model, we consider the holographic QCD confinement potential, longitudinal confinement, and a one-gluon exchange interaction with fixed coupling in the effective light front Hamiltonian. We observed a reasonable agreement of our results for the proton and $u$ quark form factors with the experimental data, however, for the Dirac form factor of $d$ quark and the neutron charge form factors deviate from the data for the basis truncation $N_{\rm max}=8$ and $K_{\rm max}=10$. We also presented the electromagnetic radii for the nucleon.   
\\
\\
{\it Acknowledgments:} CM is supported by the China Postdoctoral Science Foundation (CPSF) under the Grant No. 2017M623279 and the National Natural Science Foundation of
China (NSFC) under the Grant No. 11850410436. This work of XZ is supported by new faculty startup funding by the Institute of Modern Physics, Chinese Academy of Sciences under the Grant No.
Y632030YRC. HL is supported by the U.S. Department of Energy under Award No. DE-FG02-93ER-40762. This work of JPV is supported  by the Department of Energy under Grants Nos. DE-FG02-87ER40371, DE-SC0018223 (SciDAC4/NUCLEI), and DE-SC0015376 (DOE Topical Collaboration in Nuclear
Theory for Double-Beta Decay and Fundamental Symmetries).

\end{document}